\documentclass[3p,times]{elsarticle}

\usepackage{ecrc}

\usepackage{lineno}
\usepackage{amsmath}
\usepackage{graphicx}
\usepackage{caption}
\usepackage{subcaption}

\volume{00}

\firstpage{1}

\journalname{Nuclear Physics A}

\runauth{}

\jid{npa}

\jnltitlelogo{Nuclear Physics A}

\usepackage{amssymb}

\biboptions{square,comma,numbers,sort&compress}

\usepackage[figuresright]{rotating}

\begin{document}

\begin{frontmatter}



\dochead{}

\title{Medium-induced gluon radiation beyond the eikonal approximation}


\author[label1,label2]{Liliana Apolin\'{a}rio}
\author[label1]{N\'{e}stor Armesto}
\author[label1,label2,label3]{Guilherme Milhano}
\author[label1]{Carlos A. Salgado}
\address[label1]{Departamento de F\'{i}õsica de Part\'{i}culas and IGFAE, Universidade de Santiago de Compostela, 15706 Santiago de Compostela, Galicia-Spain}
\address[label2]{CENTRA, Instituto Superior T\'{e}cnico, Universidade de Lisboa, \\Av. Rovisco Pais, P-1049-001 Lisboa, Portugal}
\address[label3]{Physics Department, Theory Unit, CERN, CH-1211 Gen\'{e}ve 23, Switzerland}

\begin{abstract}
In this work we improve existing calculations of radiative energy loss by computing corrections that implement energy-momentum conservation, previously only implemented a posteriori, in a rigorous way. Using the path-integral formalism, we compute in-medium splittings allowing transverse motion of all particles in the emission process, thus relaxing the assumption that only the softest particle is permitted such movement. This work constitutes the extension of the computation carried out for $x \rightarrow 1$ in Phys. Lett. B718 (2012) 160-168, to all values of $x$, the momentum fraction of the energy of the parent parton carried by the emitted gluon. In order to accomplish a general description of the whole in-medium showering process, in this work we allow for arbitrary formation times for the emitted gluon (the limit of small formation times was previously employed in [J.-P. Blaizot, F. Dominguez, E. Iancu, and Y. Mehtar-Tani, JHEP1301 (2013) 143], for the $g \rightarrow gg$ splitting). We provide general expressions and their realisation in the path integral formalism within the harmonic oscillator approximation.
\end{abstract}

\begin{keyword}
Heavy-ion collisions \sep Jet Quenching \sep Jet Broadening


\end{keyword}

\end{frontmatter}

%
%

\section{Introduction}
\label{sec:intro}

\par \textit{Jet Quenching} mechanisms, the generic name given to a collection of energy loss processes that high transverse momentum objects suffer when propagating through a very hot and dense medium, are one of the best tools to analyze this new state of matter. This phenomenon has been experimentally confirmed through several observations made firstly at the Relativistic Heavy Ion Collider (RHIC) and presently at the Large Hadron Collider (LHC) (as an example, see \cite{Aamodt:2010jd,Chatrchyan:2011sx,:2012is}). Through the comparison of jet quenching models with data, it is possible to see that most models claim to describe some or all of jet quenching observables at the same time (as an example, see \cite{Apolinario:2012cg}). But even though there is a qualitative agreement with data, there is also space for improvements. The ones that will be addressed in this manuscript are related to finite energy corrections. These are particularly relevant to obtain an analytical expression that can represent the full in-medium gluon emission kinematics beyond current limitations. In this way, this new ingredient can be used as an input for Monte Carlo codes avoiding further assumptions lacking a theoretical basis. Some efforts in this direction have been already made in \cite{Apolinario:2012vy,Blaizot:2012fh}. This manuscript is organized as follows: in section \ref{sec:med}, the setup of the calculation and the results for the total spectrum are presented, including the results of the Dirac and colour structure (medium averages) in section \ref{subsec:colour}. The final conclusions are presented in section \ref{sec:conc}.

\section{Medium-induced gluon radiation}
\label{sec:med}

\par During the propagation trough an extended coloured medium, a particle can experience energy loss phenomena that can occur by medium-induced gluon radiation (the dominant mechanism for high energy particles). The interaction is mediated by very soft gluons with only transverse momenta of the order of the characteristic scale of the medium. The resulting propagator is a Green's function
\begin{align}
\label{eq:GreenFunc}
	G_{\alpha_f \alpha_i} (x_{f+}, \mathbf{x}_f; x_{i+}, \mathbf{x}_i | p_+) = & \int_{\mathbf{r} (x_{i+}) = \mathbf{x}_i}^{\mathbf{r} (x_{f+}) = \mathbf{x}_f } \mathcal{D} \mathbf{r} (\xi) \exp \left\{ \frac{ip_+}{2} \int_{x_{i+}}^{x_{f+}} d\xi \left( \frac{d \mathbf{x}}{d\xi} \right)^2 \right\} W_{\alpha_f \alpha_i} (x_{f+}, x_{i+}; \mathbf{r} (\xi) ) \, ,
\end{align}
that describes the Brownian motion of the particle in the transverse plane, from time\footnote{Using light-cone coordinates, $a=(a_0,a_x,a_y,a_z)=(a_+,a_-,{\mathbf a})$ where $a_\pm=(a_0\pm a_z)/\sqrt{2}$ and ${\mathbf a} =(a_x,a_y)$ the transverse 2-vectors.} $x_{i+}$ and initial transverse position $\mathbf{x}_i$ to time $x_{f+}$ and final transverse coordinate $\mathbf{x}_f$, at the same time that its colour field is rotated from $\alpha_i$ to $\alpha_f$ by the action of the Wilson line,
\begin{equation}
	W_{\alpha_f \alpha_i } (x_{f+}, x_{i+}; \mathbf{r} (\xi) ) = \mathcal{P} \exp \left\{ ig \int_{x_{i+}}^{x_{f+}} d\xi A_- (\xi, \mathbf{r} (\xi) ) \right\}.
\end{equation}
To describe the gluon emission off a quark in a finite medium, carrying a finite fraction $x$ of the initial energy $p$, two separate contributions must be considered: the gluon emission vertex can take place outside (eq. \eqref{eq:specout}) or inside (eq. \eqref{eq:specin}) the medium. In each case, a Green's function is associated with every in-medium propagator. This kinematical description was already made in \cite{Blaizot:2012fh} for the case of the 3-gluon vertex, but assuming small formation times, $t_{form}$, with respect to the medium length $L_+$, $t_{form} << L_+$ (infinite medium limit). While this approximation may be valid for a comfortable region of phase space, it may not hold for the partons radiated near the edge of the medium. As most Monte Carlo codes need this correction, we do not assume such constrain in the description of the in-medium showering process. The total double differential spectrum of the in-medium process where a gluon with 4-momentum $k = x p$ is emitted from a quark that remains with a 4-momentum $q = (1-x) p$, will be the sum of the two contributions:
\begin{equation}
\label{eq:spectot}
	\frac{d^2 I}{d\Omega_k d\Omega_q} = \left\langle | M_{tot} |^2 \right\rangle = \left\langle |M_{out}|^2 \right\rangle + \left\langle |M_{in}|^2 \right\rangle + 2 \text{Re} \left\langle M_{in} M^\dagger_{out} \right\rangle \, ,
\end{equation}
where $d\Omega_k = (2\pi)^{-3} d\mathbf{k} dk_+ / (2 k_+)$, and similarly for $d\Omega_q$.
Since the calculation is performed for a fixed, but arbitrary, medium configuration, an average over all possible coloured configurations must be carried out. This is represented in the above equation by $\left\langle \cdots \right\rangle$.The amplitudes for each diagram, after some algebraic simplifications, can be written as
\begin{align}
\label{eq:specout}
	\mathcal{T}_{out}  = & -\frac{g}{(2\pi)^3} \int_{-\infty}^{+\infty} d\mathbf{x} \, d\mathbf{x}_{0}\,  \text{e}^{ -i \mathbf{x} \cdot (\mathbf{k} + \mathbf{q}) + i \mathbf{x}_0 \cdot \mathbf{p}_0} \,  T^a_{B A_1} G_{A_1 A} (L_+, \mathbf{x}; x_{0+}, \mathbf{x}_0 | p_{0+}) \frac{1}{4 (k\cdot q)}  \, \bar{u} (q) \sl{\epsilon}_k^* (\sl{k} + \sl{q}) \gamma_+ \gamma_- \nonumber \\
	& \times M_h(p_{0+}) \delta (k + q - p_{0})_+ \, ,
\end{align}
\begin{align}
\label{eq:specin}
	\mathcal{T}_{in} = & \frac{ig}{(2\pi)^3} \int_{x_{0+}}^{L_+} dx_{1+} \int_{-\infty}^{+\infty} d\mathbf{x}_0 \, d\mathbf{x}_1 \, d\mathbf{y} \, d\mathbf{z} \, \text{e}^{ -i \mathbf{z} \cdot \mathbf{k} - i \mathbf{y} \cdot \mathbf{q} + i \mathbf{x}_0 \cdot \mathbf{p}_0 } G_{B B_1} (L_+, \mathbf{y}; x_{1+}, \mathbf{x}_1 | q_+)  T_{B_1 A_1}^{a_1} G_{A_1 A} (x_{1+}, \mathbf{x}_1; x_{0+} \mathbf{x}_0 | p_{0+})  \nonumber \\
	&\times G_{aa_1} (L_+, \mathbf{z}; x_{1+}, \mathbf{x}_1| k_+) \frac{1}{2} \bar{u}(q) \sl{\epsilon}_k^* \gamma_- M_h(p_{0+}) \delta (k + q - p_0)_+\, ,
\end{align}
where $\left\langle |M_{out}|^2 \right\rangle = \frac{\left\langle |\mathcal{T}_{out}|^2 \right\rangle}{\sigma_{el}}$, and $\sigma_{el}$ is the cross section of the in-medium elastic channel. The fundamental coloured indices are represented by uppercase Latin letters and the adjoint ones by lowercase Latin letters. The coordinate $x_{0+}$ refers to the beginning of the medium, $x_{1+}$ is the longitudinal position of the in-medium emission vertex, and $\mathbf{x}_0$, $\mathbf{x}_1$, $\mathbf{y}$ and $\mathbf{z}$ are the transverse coordinates of the corresponding propagators at times $x_{0+}$, $x_{1+}$ and $L_+$ respectively. The $M_h(p_{0+})$ represents the hard scattering amplitude that generates the original quark and is assumed to be unmodified by the medium. 

\subsection{Computing the medium averages}
\label{subsec:colour}

\par Due to the high energy limit that is assumed throughout this calculation, the medium averages can be computed locally, i.e., it is understood that the dynamics that lead to the modifications of the medium colour structure occur in a timescale that is much larger than the propagation time of the penetrating particle. This fact, together with the \textit{decoupling} of the Brownian propagator from the Wilson line in equation \eqref{eq:GreenFunc}, allows us to perform a diagrammatic separation into regions where there is a fixed number of Wilson lines. Consequently, it is possible to solve independently the colour structure and the transverse momentum broadening. 

\begin{figure}[htbp]
\centering
	\includegraphics[width=0.6\textwidth]{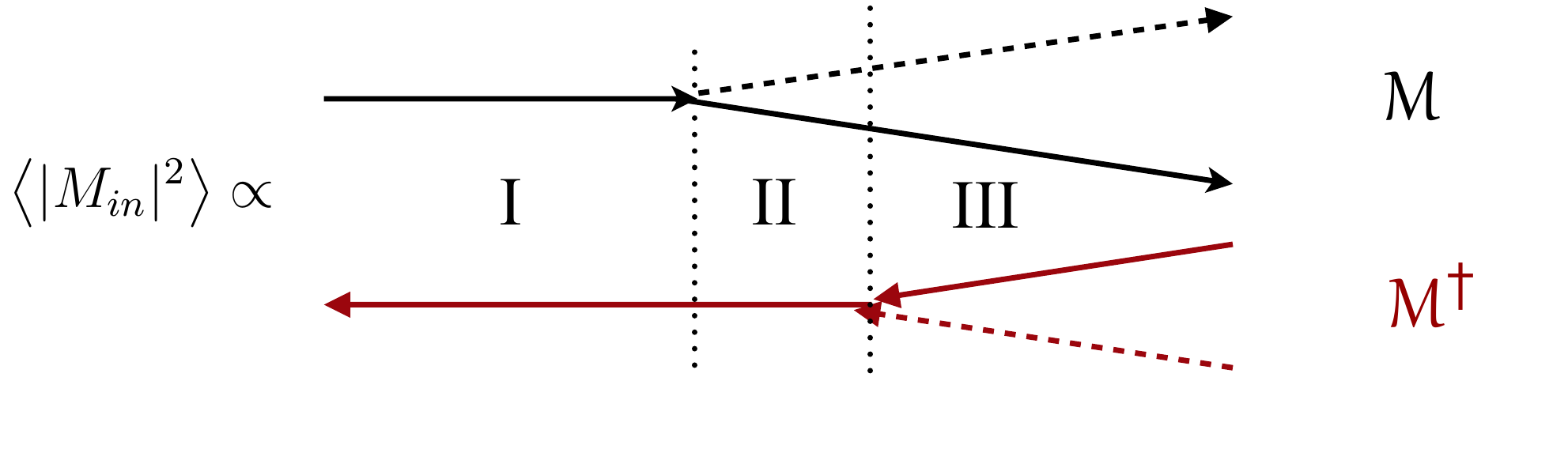}
	\caption{Diagrammatic separation of the $\left\langle |M_{in}|^2 \right\rangle$, where the amplitude is represented by black lines and the conjugate one by red lines. The solid lines represent the quarks while the dashed ones the gluons.}
	\label{fig:Minin}
\end{figure}

\par As we do not assume constrains for the $t_{form}$ three regions are identified, being all non-trivial contributions only present in the $\left\langle |M_{in}|^2 \right\rangle$ term (see figure \ref{fig:Minin}). To calculate the medium averages in each region, we assume that the particle undergoes multiple soft scatterings with the medium. To simplify the calculation of the several $n$-point functions from each region, we work in the large $N_c$ limit, and thus, the gluon can be understood as a quark-antiquark pair. The procedure to calculate the medium averages is to expand the Wilson lines in an infinitesimal interval,
\begin{equation}
	W (\mathbf{x})_{ij} = W (\mathbf{x})_{i\alpha} \left\{ \delta_{i \alpha} \left( 1 - \frac{C_F}{2} B(0) \right) - i T^a_{\alpha j} A^a (\mathbf{x}) \right\} \, ,
\end{equation}
up to second order of the fields, $B(\mathbf{x} - \mathbf{y}) \propto \left\langle A^a (\mathbf{x}) A^a (\mathbf{y}) \right\rangle$, re-iterate the process until all the longitudinal path is covered and the result can be exponentiated afterwards. Doing so, the simplest object that one can calculate, the dipole formed by two Wilson lines (2-point function), reads to\footnote{The colour pre-factor $C_F$ can be part of the $\sigma$ definition.}:
\begin{equation}
\label{eq:dipole}
	\frac{1}{N} \text{Tr} \left\langle W(\mathbf{x}) W^\dagger(\mathbf{y}) \right\rangle = \text{e}^{ C_F v(\mathbf{x} - \mathbf{y})} = \exp \left\{ - \frac{C_F}{2} \int dx_+ \sigma(\mathbf{x} - \mathbf{y}) n(x_+) \right\} \, ,
\end{equation}
where $v(\mathbf{x} - \mathbf{y}) = B(\mathbf{0}) - B(\mathbf{x} - \mathbf{y})$, $\sigma (\mathbf{x} - \mathbf{y})$ is the dipole cross section and $n(x_+)$ the longitudinal density of scattering centers. Applying this procedure, region I is trivially given by a dipole (equation \eqref{eq:dipole}) plus a Dirac delta function that allows to close region II into a quadrupole structure, that factorizes into two dipoles in the large $N_c$ limit:
\begin{align}
	& \left\langle \left[ W (\mathbf{r}_q) W^\dagger (\mathbf{r}_g) \right]_{ij} \left[ W (\mathbf{r}_g) W^\dagger (\mathbf{r}_q) \right] _{kl} \right\rangle \xrightarrow[N_c \rightarrow \infty]{} \left\langle  W (\mathbf{r}_q) W^\dagger (\mathbf{r}_g) \right\rangle \left\langle W (\mathbf{r}_g) W^\dagger (\mathbf{r}_{\bar{q}}) \right\rangle \delta_{ij} \delta_{kl} \, .
\end{align}
The transverse coordinate of the quark in the (complex conjugate) amplitude is denoted by $\mathbf{r}_{q(\bar{q})}$ and the one from the gluon by $\mathbf{r}_g$. Finally, region III, in the large $N_c$ limit, closes as an independent dipole and quadrupole:
\begin{align}
\label{eq:sextupole}
	& \left\langle \text{Tr} \left( W^\dagger (\mathbf{r}_{\bar{g}}) W (\mathbf{r}_g) \right) \text{Tr} \left( W^\dagger (\mathbf{r}_g) W (\mathbf{r}_{\bar{g}}) W^\dagger (\mathbf{r}_{\bar{q}}) W (\mathbf{r}_q) \right)  \right\rangle \nonumber \\
	& \xrightarrow[N_c \rightarrow \infty]{} \left\langle \text{Tr} \left( W^\dagger (\mathbf{r}_{\bar{g}}) W (\mathbf{r}_g) \right) \right\rangle \left\langle \text{Tr} \left( W^\dagger (\mathbf{r}_g) W (\mathbf{r}_{\bar{g}}) W^\dagger (\mathbf{r}_{\bar{q}}) W (\mathbf{r}_q) \right)  \right\rangle \, .
\end{align}

\par
\begin{figure}[htbp]
\centering
	\includegraphics[width=0.8\textwidth]{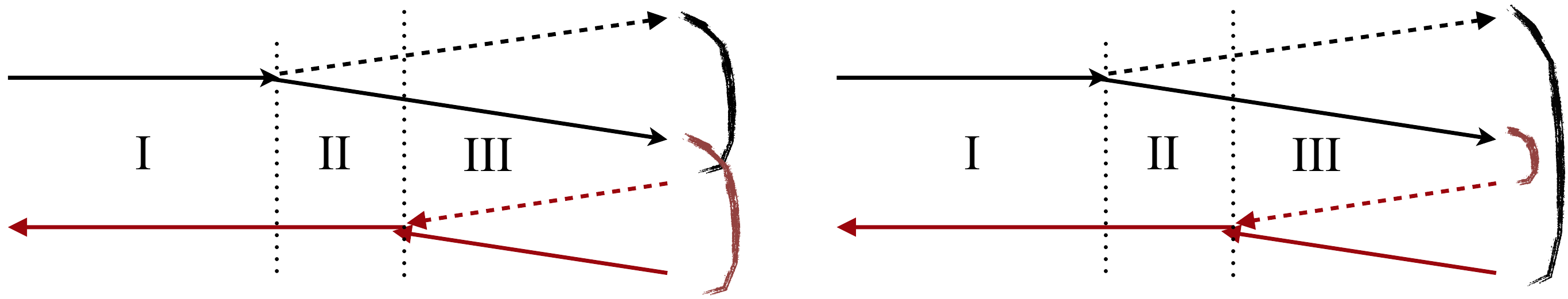}
	\caption{Diagrammatic representation of the two propagation regimes found with equation \eqref{eq:quadripole}. Region I translates the random walk of the initial quark, region II the gluon formation time, during which quark and gluon are colour correlated by definition, and finally, in region III (left) quark and gluon propagate independently or (right) remain colour connected.}
	\label{fig:result}
\end{figure}

 It can be shown that, within the employed approximations, where the dipole cross section is approximated by its small distance component\footnote{It is possible to identify the transport coefficient $\hat{q}$ that translates the average transverse momentum squared acquired by the particle when crossing the medium per mean free path, $\lambda$.} \cite{Zakharov:1996fv}, $\sigma (\mathbf{r}) = \frac{1}{2} \hat{q} \mathbf{r}^2$, the 4-point function from the above equation factorizes into a linear combination of two dipoles with pre-factors $A_1$ and $A_2$:
\begin{align}
\label{eq:quadripole}
	& \left\langle \text{Tr} \left( W^\dagger (\mathbf{r}_g) W (\mathbf{r}_{\bar{g}}) W^\dagger (\mathbf{r}_{\bar{q}}) W (\mathbf{r}_q) \right)  \right\rangle \propto \nonumber \\
	& A_1 \left\langle \text{Tr} \left( W^\dagger (\mathbf{r}_g) W (\mathbf{r}_{\bar{g}}) \right) \right\rangle \left\langle \text{Tr} \left( W^\dagger (\mathbf{r}_{\bar{q}}) W (\mathbf{r}_q) \right)  \right\rangle -  A_2 \left\langle \text{Tr} \left( W^\dagger (\mathbf{r}_g) W (\mathbf{r}_{q}) \right) \right\rangle \left\langle \text{Tr} \left( W^\dagger (\mathbf{r}_{\bar{q}}) W (\mathbf{r}_{\bar{g}}) \right)  \right\rangle .
\end{align}
The first term, schematically represented in figure \ref{fig:result} (left), is dominated by an independent propagation with the factorization of both final particles, a contribution that was already identified in \cite{Blaizot:2012fh}. In this case, the final particles will experience broadening independently. In the second term both particles are correlated and therefore, will emit coherently (see figure \ref{fig:result}, right). 

\par The soft \cite{Zakharov:1996fv,Baier:1996sk,Salgado:2003gb,Gyulassy:2000er} and hard \cite{Apolinario:2012vy} limit radiation spectra are recovered from equation \eqref{eq:quadripole} in the considered limits, i.e., when $\mathbf{r}_q = \mathbf{r}_{\bar{q}}$ and $\mathbf{r}_g = \mathbf{r}_{\bar{g}}$, respectively. The result is the dipole formed by the final gluon or quark in each case.

\section{Conclusions}
\label{sec:conc}

\par In this work we were able to extend previous derivations of the in-medium gluon emission off a quark beyond the eikonal approximation, for the case of a double differential spectrum, in energy and transverse momentum (finite energy corrections to the inclusive energy spectrum were already calculated in \cite{Zakharov:1996fv}). In particular, we were able to assign a Brownian motion in the transverse plane to all propagating particles, going beyond our previous work of \cite{Apolinario:2012vy}, and as previously done in \cite{Blaizot:2012fh}. Nonetheless, although the results are presented in the large $N_c$ limit, the results are improved with respect to \cite{Blaizot:2012fh} as we do not assume any constrain on the $t_{form}$. Still, the results previously derived are recovered in the considered limit. Moreover, since a finite medium is considered, the vacuum interference term is included in the calculation of the Dirac and colour structure. 

\small{\textbf{Acknowledgments:} This work was supported by the European Research Council grant HotLHC ERC-2011-StG-279579, and by Funda\c{c}\~{a}o para a Ci\^{e}ncia e a Tecnologia of Portugal under projects  CERN/FP/123596/2011 and SFRH/BD/64543/2009}





\bibliographystyle{elsarticle-num}
\bibliography{Bibliography}







\end{document}